\documentclass{article}

\usepackage{graphicx}
\usepackage{amsmath,amssymb,bm}
\usepackage{float,times}

\begin{document}

\title{Dust and gas emission from cometary nuclei: the case of comet 67P/Churyumov-Gerasimenko}

\author{\small
Tobias~Kramer\textsuperscript{a,b}$^{\ast}$\thanks{$^\ast$Corresponding author. Email: kramer@zib.de}, Matthias~Noack\textsuperscript{a}, Daniel~Baum\textsuperscript{a}, Hans-Christian~Hege\textsuperscript{a}, Eric~J.~Heller\textsuperscript{b}\\\small
\textsuperscript{a}Zuse Institute Berlin (ZIB),
 Takustr. 7, 14195 Berlin, Germany\\\small
\textsuperscript{b}Department of Physics, Harvard University,
12 Oxford St, Cambridge, MA 02138, U.S.A.
}
\date{\small \today\\
Advances in Physics X (2018)
doi: 10.1080/23746149.2017.1404436
}

\maketitle

\begin{abstract}
Comets display with decreasing solar distance an increased emission of gas and dust particles, leading to the formation of the coma and tail.
Spacecraft missions provide insight in the temporal and spatial variations of the dust and gas sources located on the cometary nucleus.
For the case of comet 67P/Churyumov-Gerasimenko (67P/C-G), the long-term observations from the Rosetta mission point to a homogeneous dust emission across the entire illuminated surface.
Despite the homogeneous initial distribution, a collimation in jet-like structures becomes visible.
We propose that this observation is linked directly to the complex shape of the nucleus and projects concave topographical features into the dust coma.
To test this hypothesis, we put forward a gas-dust description of 67P/C-G, where gravitational and gas forces are accurately determined from the surface mesh and the rotation of the nucleus is fully incorporated.
The emerging jet-like structures persist for a wide range of gas-dust interactions and show a dust velocity dependent bending.
\end{abstract}

\section{Introduction}

The modeling of the dust and gas emissions of cometary nuclei requires linking physical and chemical processes ranging from the formation and release of molecules to the emission of macroscopic particles, driven by solar radiation and gravitational forces.
Advanced models relating gas emission and subsequent acceleration of dust grains from the surface layer of a comet have been developed over the years based on dust Monte-Carlo simulation (DMC) techniques \cite{Zakharov2009,Combi2012}.
Space-craft observations starting with the Halley flyby 1986 reveal complex jet/filament structures near the cometary nucleus \cite{Keller1994}.
The Rosetta mission \cite{Schulz2009} has followed from close distance (about 50-200~km) the evolution of 67P/C-G over more than 800 days of its orbital period of 2400 days and provides the most comprehensive data set and measurements of the space environment around a comet.
The multitude of observations and large data sets enters the parametrization and reconstruction of the gas and dust environment, which face various obstacles and uncertainties.
While the gas measurements have been performed directly at the space-craft position with a high temporal resolution (up to every minute by the Cometary Pressure Sensor (COPS) \cite{Balsiger2007}), other instruments recorded data less frequently and have larger instrumental uncertainties.
In particular, the direct measurement of the local dust density is affected by charge on the dust and the spacecraft \cite{Fulle2015}.
The recorded compact dust particles originating from 67P/C-G have sizes ranging from $0.03$ to $1$~mm, while fluffy dust aggregates (sizes $0.2$ to $2.5$~mm) have been registered with the Grain Impact and Dust Analyzer (GIADA) instrument \cite{Fulle2015}. 
The typical dust velocities fall in the range $2$ to $6$~m/s from August~2014 to January~2015 at a distance from the nucleus of $\sim 20$ to $50$~km.

The prominent dust tail of comets displays a non-homogeneous and non-spherical structure at various length scales, ranging from few kilometers to $10^5$~km for 67P/C-G.
The large-scale structure of the tail as seen from Earth bound telescopes is in addition shaped by the radiation pressure and the solar wind.

Here, we focus on the origin of the innermost ($<20$~km) jet-like structure observed by the OSIRIS Wide Angle Camera (WAC) on Rosetta \cite{Keller2007}, which is not accessible with Earth-bound telescopes.
The scattering of light by dust grains is wavelength dependent.
To resolve dust and gas near the nucleus, the OSIRIS WAC camera is equipped with different narrow bandwidth filters, which require longer exposure times \cite{Keller2007}.
Before the Rosetta observations, it was suggested that the collimation of dust particles into jets is linked to surface structures by either introducing isolated areas of cometary activity with respect to gas emission (``spots'') \cite{Combi2012}, topographical features such as pits and floors of craters \cite{Keller1994}, or terraced regions \cite{Farnham2013}.
For 67P/C-G, cliffs were proposed as sources of dust jets \cite{Vincent2015a}.
In Sect.~\ref{sec:model} we study a very generic collimation and dust-jet formation mechanism based on the assumption of a homogeneous surface activity, without restricting it to few active areas.
Collimation and formation of jets in the homogeneous activity model result either from the convergence of trajectories originating from a concave area, or are the result of bending of trajectory bundles due to the Coriolis effect and gravity.
We start from a best-fit gas density and velocity field obtained from an inverse model of the COPS measurements in April 2015 \cite{Kramer2017}, described in Sect.~\ref{sec:gas}.
In the second step (Sect.~\ref{sec:dust}), we consider the emission of homogeneously distributed dust grains from the cometary surface and demonstrate a mechanism for collimated dust jets despite uniform emission.
The shown collimation model is robust against variations in gas-dust interaction, but selective with respect to the particle size.
We fully incorporate the rotation of the nucleus and show that the Coriolis effect puts strong constraints on dust velocities consistent with observations of curved jets from 67P/C-G (Sect.~\ref{sec:coriolis}).
In Sect.~\ref{sec:results} we test the model for a series of varying positions of the Rosetta spacecraft with respect to the nucleus and compare the observed images with column densities from the model.
Jet-like structures are reproduced at locations seen in Rosetta images and the comparison of model and observation allows one to determine the dust velocity in addition to verify the proposed collimation mechanism.
A summary and open questions regarding the initial gas-dust interaction are discussed in the conclusions.

\begin{table}[t]
\begin{center}
$\begin{array}{lrl}
\text{mass}    & 1.0\times 10^{13} &\text{kg}\\
\text{volume} & 18.8  & \text{km}^3 \\
\text{density} & 532 &\text{kg/m}^{3}\\
\text{rotation period} & 12.3-12.4 &\text{h} \\
\text{rotation axis right ascension} \;\alpha &69.6&{}^\circ  \\\text{rotation axis declination}     \;\delta &64.0&{}^\circ 
\end{array}$
\end{center}
\caption{\label{tab:cometprop} Properties of the model for 67P/C-G given in Tab.~4 of Ref.~\cite{Jorda2016}}
\end{table}

\section{Homogeneous surface-activity model of the nucleus}\label{sec:model}

Modelling radiation-driven activity of comets requires setting up a three-dimens\-ional model of the physical (temperature, density/porosity, gas/solid phase) and chemical properties (molecular reactions) throughout the nucleus \cite{Huebner2006}.
In computational fluid dynamics approaches, the nucleus is subdivided into volumetric cells and the diffusion and heat equations are solved \cite{Huebner2006,BruckSyal2013}.
The dust is subsequently embedded into the gas-model and accelerated by gas drag within the porous mantle and outside the cometary body.
Gas emission and dust acceleration were previously considered based on different physical processes and dust origins.
Refs.\ \cite{Yelle2004,Huebner2006} consider dust acceleration due to gas drag already in the outer porous mantle of the comet within a few meters below the surface.
Upon leaving the mantle and entering the tenuous atmosphere of the comet, the gas density drops sharply and the subsequent dust motion is considered to be ballistic, \cite{Yelle2004,BruckSyal2013}.
Ref.~\cite{Combi2012} starts dust grains initially resting at the surface of a spherical non-rotating comet and takes the gas drag outside the surface of the nucleus as the only accelerating effect.
This results in dust velocities of $\sim 10-300$~m/s at 10~km distance.
As shown in Sect.~\ref{sec:coriolis}, these larger than observed velocities are partly a consequence of neglecting the rotation of the nucleus, leading to a divergence of gas and dust velocity directions.
The gas eventually disperses in space and the gas density falls of as $1/r^2$ with increasing distance $r$, while the solid dust grains remain compact particles under the influence of external forces such as gravitation and radiation pressure.
For the homogeneous activity model, we start by computing the gas density and gas velocity field around the nucleus and subsequently consider the dust motion under gas drag.

\subsection{Shape model}\label{sec:shape}

To obtain a homogeneous sampling of the comet with respect to the number of emitted particles (either dust or gas molecules) per surface area, correctly weighting Monte-Carlo methods can be used or alternatively a surface mesh with equal-area mesh cells.
We employ the latter approach due to its computational advantages for the efficient parallelization of the trajectory computation and subsequent generation of dust density maps.

\begin{figure}
\begin{center}
\includegraphics[width=\textwidth]{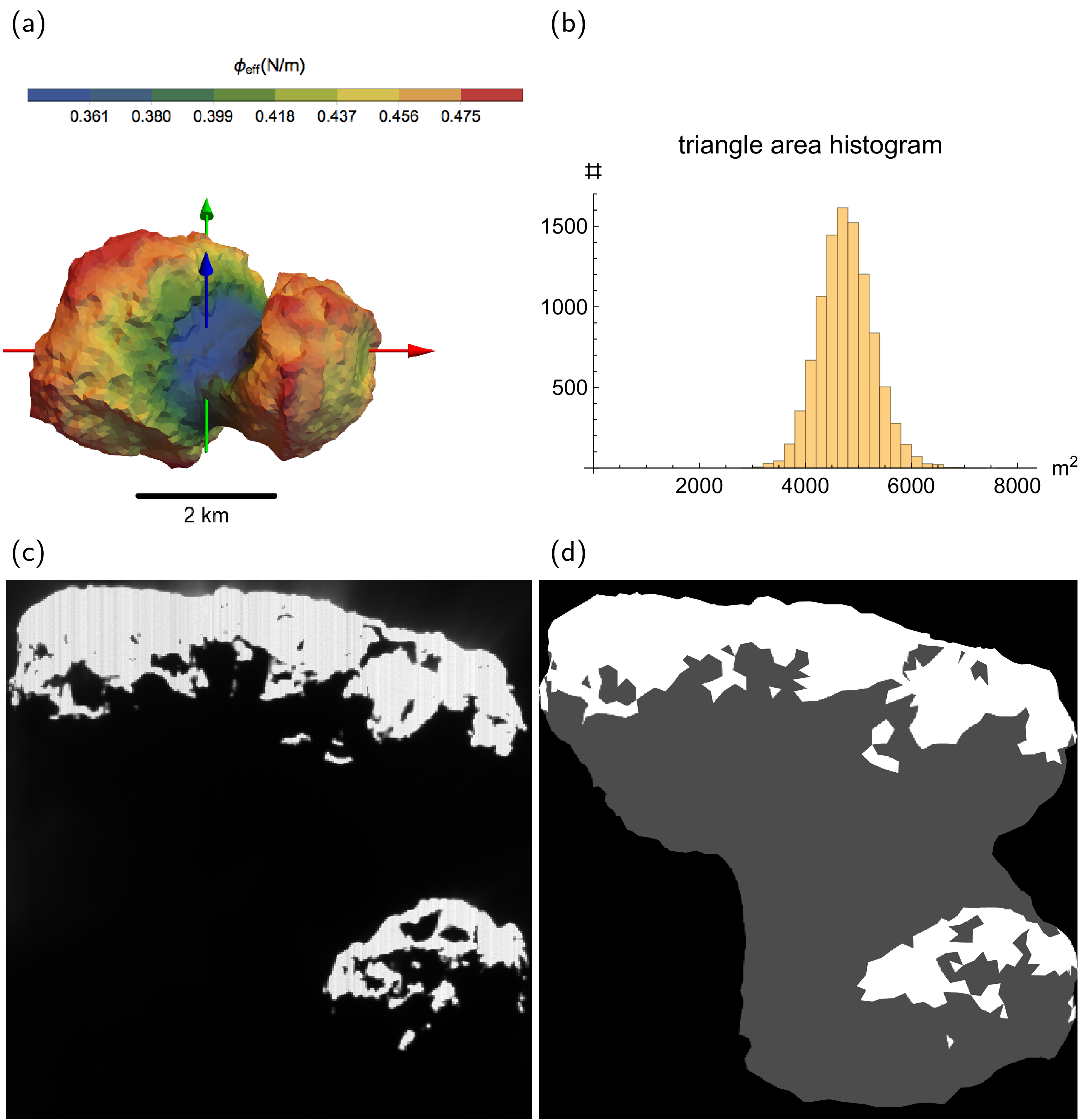}
\end{center}
\caption{\label{fig:mesh} (a) Parametrization of the shape of comet 67P/C-G by 9996 triangles;
color denotes the local effective potential $\phi_\text{eff}$ (gravitation and centrifugal contribution); 
the blue arrow along $\hat{e}_z$ indicates the axis of rotation.
(b) Histogram of the triangle area of the surface triangulation.
(c) OSIRIS WAC image (time: 2014-04-22 21:08 UTC, id: W20150422T210738516ID30F18, filters: empty+VIS610, exposure: 7.8~s, direction to Sun upwards), with contrast adjusted to emphasize sunlit and shadowed areas.
(d) Synthetically simulated image of the nucleus from Rosetta's viewpoint with illuminated areas shown in white and shadowed areas in grey using the 9996 triangles shape model. 
}
\end{figure}

Our starting point is the surface mesh of the comet with detailed three-dimensional topography, including concave areas.
The shape model is derived from the higher resolution model obtained by Mattias~Malmer \cite{MalmerShapeModelNovmeber2015} based on Rosetta navigation camera (NAVCAM) pictures.
The shape model is uniformly remeshed using the {\em Surface Remesh} module \cite{ZilskeLameckerZachow2008,surazhsky2003explicit} of the {\em Amira} software \cite{Hansen2004}. 
The resulting mesh in Fig.~\ref{fig:mesh}a contains 9996 triangles with mean area $4772$~m$^2$.
The key property of the re-meshed surface is the small spread of the areas of the mesh cells shown in the histogram (Fig.~\ref{fig:mesh}b).
This allows one to simulate a homogeneous surface emission rate by assigning one representative test particle to each cell with well-defined surface normal direction and uniform areal coverage.
Fig.~\ref{fig:mesh}d shows a synthetic view of the illuminated shape model with sunlit areas shown in white. 
The actual OSIRIS image from the same viewpoint and date is shown in Fig.~\ref{fig:mesh}c and confirms the quality of the shape model.

\subsection{Gas model}\label{sec:gas}

\begin{figure}[t]
\begin{center}
\includegraphics[width=\textwidth]{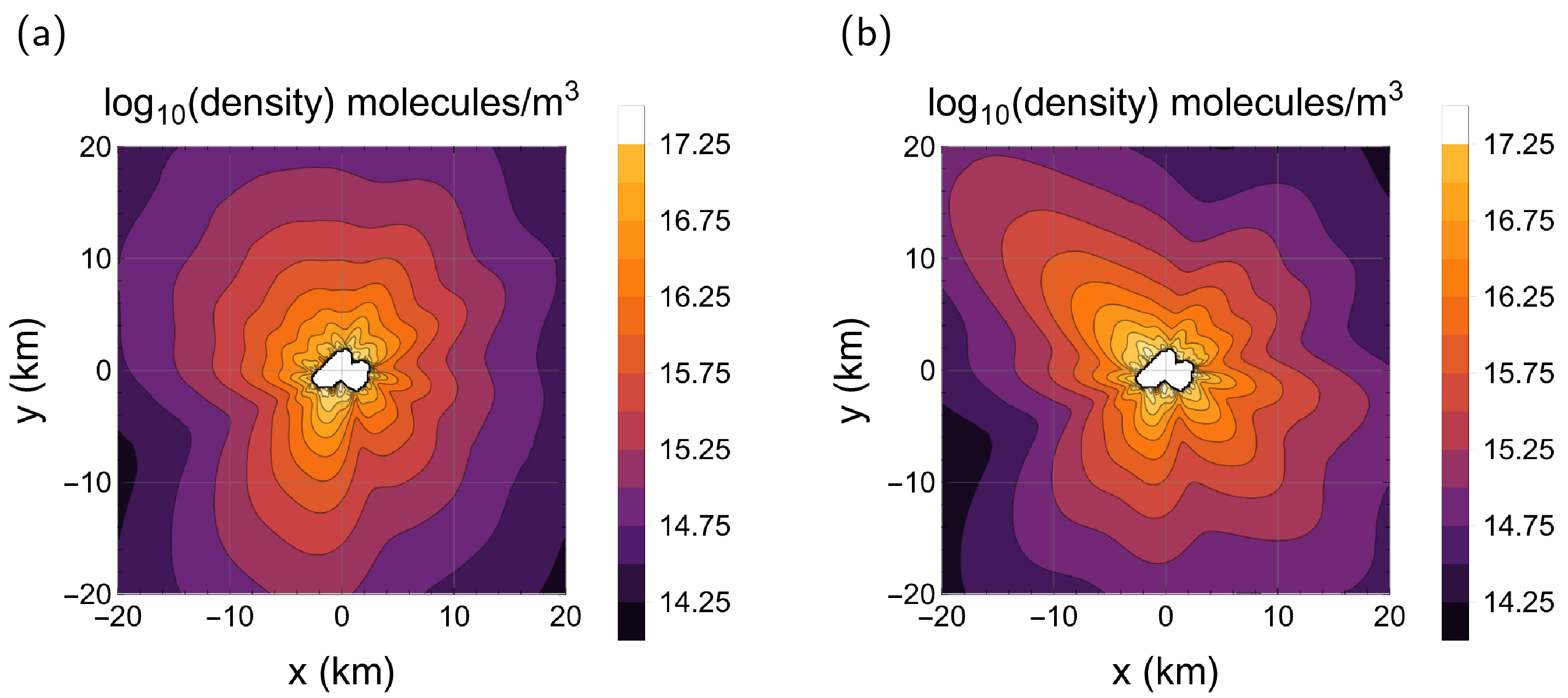}
\end{center}
\caption{\label{fig:gasdensitymap} 
Gas number density $N_\text{gas}$ around comet 67P/C-G (slice at $z=0$)
for (a) a best fit coma model for April 2015 to the COPS data set \cite{Kramer2017} and for (b) homogeneous gas emission.}
\end{figure}

The gas velocity $v_\text{gas}$ is taken as the root mean squared speed of the Maxwellian distribution (\cite{Vincent2015a}, eq.~(3))
\begin{equation}\label{eq:gasvel}
\tilde{v}_{\rm th}=\sqrt{\frac{3 k_B T}{m_\text{gas}}},
\end{equation}
with the Boltzmann constant $k_B$ and the mass of the gas molecule $m_\text{gas}$
resulting in $\tilde{v}_{\rm th}\approx 540$~m/s for H$_2$O molecules at the temperature $T=210$~K.
The dust collimation does not depend on the precise value of the gas velocity.
We take the reconstructed gas coma from Ref.~\cite{Kramer2017} for the month of April 2015, which reproduces the 21125~COPS gas measurements with an error below 12~percent.
The number of COPS measurements restricts the resolution of the shape model, since a converged inversion of the measured data requires to take about twice as many observations compared to the potential gas emitting sources located on the shape model.
In the following we use the 9996 gas sources distributed on the shape model introduced in the previous section and considered in Ref.~\cite{Kramer2017}.
A slice through the gas density is shown in Fig.~\ref{fig:gasdensitymap}a. The gas density close to the comet approaches $10^{17}$~H$_2$O molecules/m$^3$, and drops off to values $\sim 10^{13}$~H$_2$O molecules/m$^3$ at typical Rosetta distances of $150$~km.
The gas coma is assumed to be static in time for the following simulation with an illumination-dependent dust lift-off condition introduced later.

\subsection{Dust dynamics}\label{sec:dust}

In the next computational step, we obtain the dust dynamics within the expanding gas cloud. 
The positions and velocities of representative dust particles are specified in the body-fixed (=rotating) frame, while the comet rotates with period $T_\text{rot}=\frac{2\pi}{\omega}$ in the inertial frame around the $\hat{e}_z$-axis (see Fig.~\ref{fig:mesh}).
The gravitational potential $\phi$ of the nucleus is calculated under the assumption of a homogeneous mass distribution from the polyhedral representation \cite{Conway2014}.
The acceleration on a dust particle embedded in the more rapidly moving gas is given by:
\begin{eqnarray}\label{eq:accdust}
{\bm a}_{\text{dust}}({\bm r})
&=&{\bm a}_\text{gas drag}+{\bm a}_\text{grav}+{\bm a}_\text{centrifugal}+{\bm a}_\text{Coriolis}\\\nonumber
&=&\frac{1}{2}C_d \alpha N_\text{gas}  ({\bm r}) m_\text{gas} ({\bm v}_\text{gas}-{\bm v}_\text{dust})|{\bm v}_\text{gas}-{\bm v}_\text{dust}|
- \nabla \phi({\bm r})\\\nonumber
&&-2{\bm \omega}\times {\bm v}_\text{dust}-{\bm \omega}\times({\bm \omega}\times{\bm r}).
\end{eqnarray}
The initial direction of the velocity $v_{\rm init}$ of the dust is taken to be at distance $h_{\rm init}$ from the gas source at the center of the mesh cell along the outward surface normal $\hat{{\bm n}}_i$ of the mesh-cell $i$:
\begin{eqnarray}
{\bm v}_i(t=0)&=&v_{\rm init}\;\hat{{\bm n}}_i,\quad i=1,\ldots,9996\\
{\bm r}_i(t=0)&=&(\text{center of triangle}(i))+h_{\rm init}\; \hat{\bm n}_i.
\end{eqnarray}
The gas-drag acceleration depends on the dust-particle mass $m_\text{dust}$ and radius $R_\text{dust}$, as well as on the momentum of the gas molecules.
In the following we set $C_d=2$ (spherical particle, \cite{Keller1994}) and vary the parameter $\alpha$ related to the dust particle radius $R_\text{dust}$ and density $\rho_\text{dust}$
\begin{equation}\label{eq:alphagas}
\alpha=\frac{3}{4}\frac{1}{\rho_\text{dust} R_\text{dust}}
\end{equation}
from $4$ to $16$. 
For a density of $\rho_\text{dust}=440$~kg/m$^3$ \cite{Marschall2015}, this corresponds to particles sizes $R_\text{dust}=426$~$\mu$m to $106$~$\mu$m.

In addition to the gas drag encountered by dust outside the nucleus, different scenarios for the initial acceleration of dust already close to the surface of the nucleus were considered in Ref.~\cite{Huebner2006}, chapter~3.4, 
which models dust flow and acceleration in the pores of the mantle. 
In this case, dust is ejected out of the nucleus with finite initial velocity, gained within less than a meter distance from the surface.
A large initial dust acceleration is part of most direct model-to-observation comparisons, since a completely smoothly distributed gas outflow across the entire surface facets results in an insufficient gas drag near the surface \cite{Vincent2015a}.
In the present model, the distribution of the gas emitters buried beneath the center of each triangular face leads to a focused, cone-shaped gas emission \cite{Kramer2017}, which efficiently accelerates the dust close to the surface.
For the dust we set $v_{\rm init}=0$~m/s and take the distance of a representative dust tracer particle to the point-like gas source of $h_{\rm init}=50$~m, which is not to be taken as a physical distance, but parametrizes a realistic aperture of the gas source compared to the singular $\delta$-distribution of a mathematical point source.
One direct consequence of the fast initial velocity gain is that the local topography of the cometary surface is imprinted on the initial velocity direction, which is then expected to be (on average) along the surface normal.

\section{Results for comet 67P/Churyumov-Gerasimenko}\label{sec:rosetta}

\begin{figure}[t]
\begin{center}
\includegraphics[width=\textwidth]{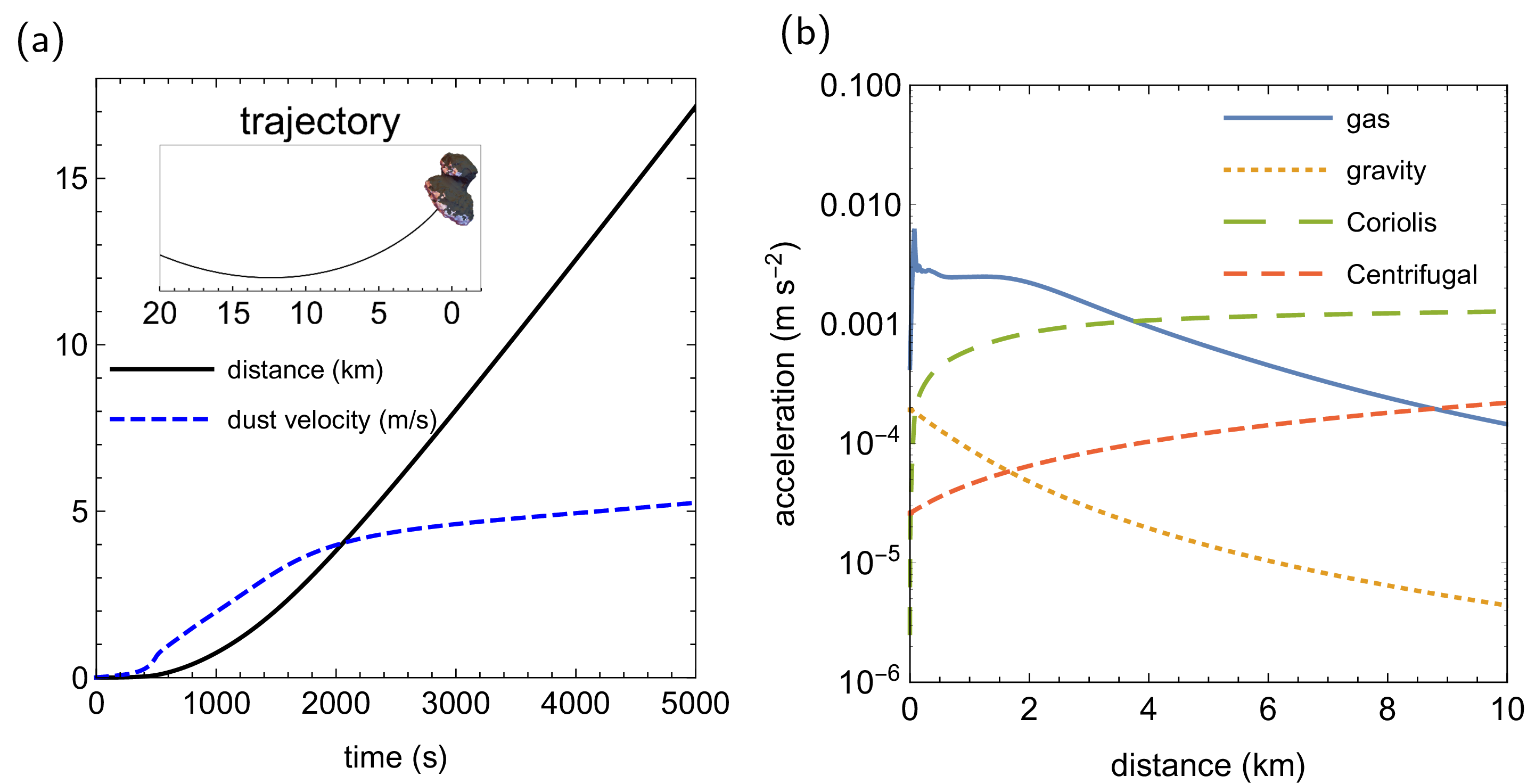}
\end{center}
\caption{\label{fig:forces} 
(a) Velocity and distance of a dust grain ($\alpha=16$) emanated from the large lobe (see inset for the trajectory) within the first 5000~s.
(b) Accelerating forces acting on the particle, divided into gas drag, Coriolis effect (acting orthogonal to the velocity), gravitation, and centrifugal components.
}
\end{figure}

It is instructive to study the relative contribution of the different terms contained in Eq.~(\ref{eq:accdust}) for the specific case of 67P/C-G.
As exemplary particle, we consider a dust grain emitted from the almost flat surface of the big lobe of the comet moving in the gas field shown in  Fig.~\ref{fig:gasdensitymap}a.
The velocity and position of the dust particle are determined by the gas drag, the Coriolis effect, gravitational and centrifugal forces, which vary along the trajectory (Fig.~\ref{fig:forces}).
While the influence of gas drag drops rapidly with increasing distance, the rotation of the comet has a large influence on the dust  trajectories through the velocity-dependent Coriolis effect.
For 67P/C-G the rotation of the nucleus cannot be neglected as was done in previous studies \cite{Zakharov2009}.
If only the Coriolis force is present, the motion of a particle with its velocity vector pointing radially outward in the plane orthogonal to the rotation axis follows a circle with radius
\begin{equation}\label{eq:rcor}
R_\text{Coriolis}=\frac{|{\bm v}_\text{dust}({\bm r})|}{2\omega}.
\end{equation}
The Rosetta observations allow one to determine $R_\text{Coriolis}$ and to estimate the dust velocity, see Sect.~\ref{sec:coriolis}.

\subsection{Sunlight driven activity model}

\begin{figure}[t]
\begin{center}
\includegraphics[width=\textwidth]{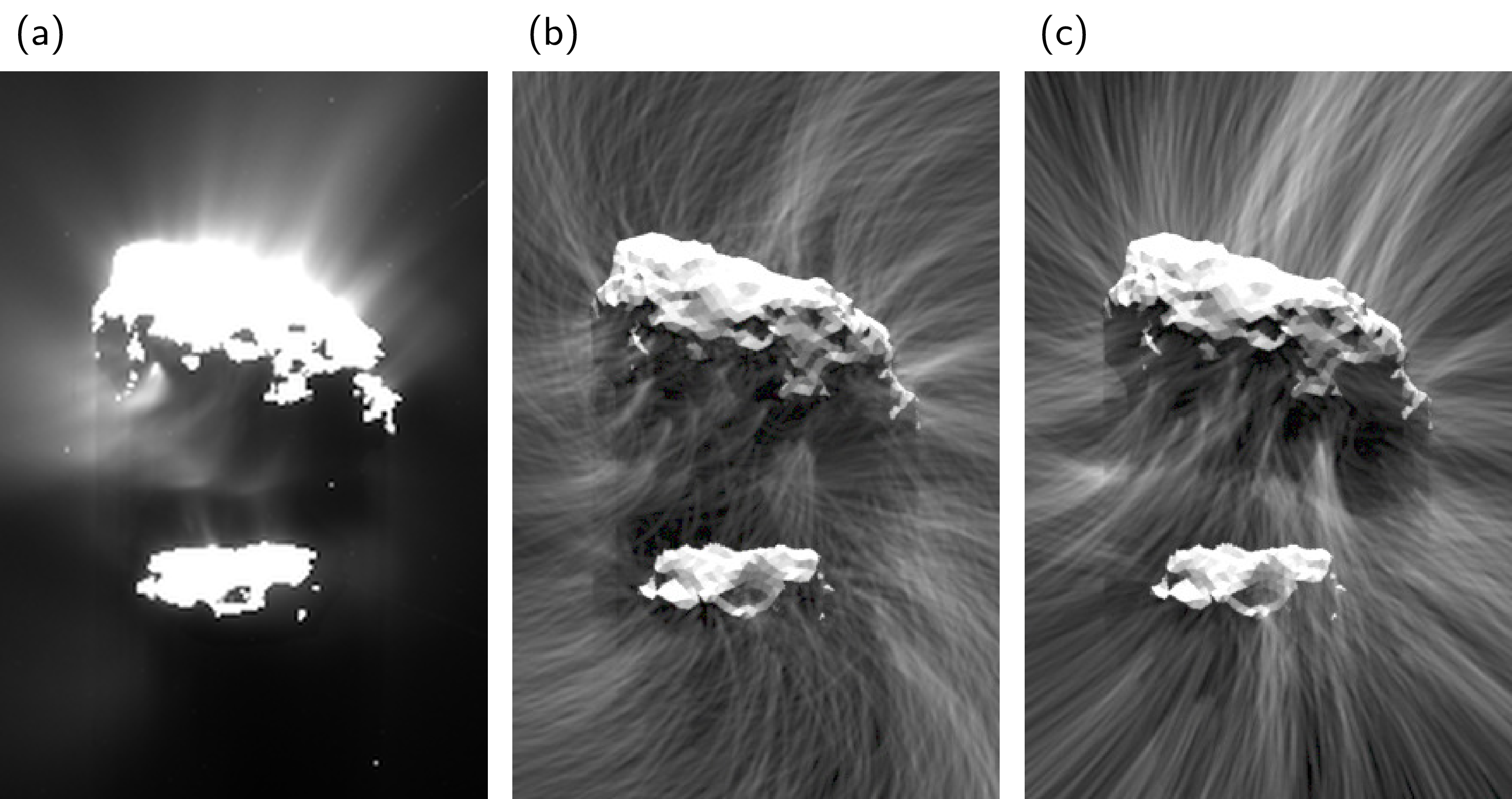}
\end{center}
\caption{\label{fig:night}
Parametric study of varying dust-gas interactions ($\alpha=0.5$-$1$,$8$) for the Rosetta viewpoint on 2015-04-24 09:30 UTC (a), contrast enhanced WAC image 
(id: W20150424T092929750ID30F13, filters: empty+UV375, exposure: 36.45~s, direction to Sun upwards) to emphasize the dust in the shadowed neck area.
Best agreement of the OSIRIS image with the simulation is obtained for small $\alpha=0.5$-$1$ in the shadowed area (b), while outside the nucleus dust from sunlit areas fits better with higher $\alpha=8$ (c).
All 9996 dust trajectories are included and illuminated by ambient lighting.
}
\end{figure}

A direct comparison with Rosetta observations requires considering the solar illumination condition and spacecraft position with respect to the comet.
Within the homogeneous activity model, the only adjustable parameter is the dust-gas interaction parameter $\alpha$, which translates to the dust grain size, see eq.~(\ref{eq:alphagas}).
We construct a directional sunlight model using the sun-comet vector obtained from the HORIZON ephemeris file of the heliocentric position of the comet.
The shape model is then rotated to align the rotation axis with the pole location in Tab.~\ref{tab:cometprop}.
Any precession of the rotation axis or change in rotation period due to torque acting on the comet is neglected.
For the given time of observation we determine Rosetta's viewpoint and the sub-solar point from the archived SPICE (Spacecraft, Planet, Instrument, Camera, Events) data of the Rosetta mission and compute the solar illumination on the surface.
We restrict the dust emitting sources with a sunlight-driven activity model and remove all dust trajectories that originate from shadowed surface cells.
This simplified model neglects any decline of the gas density at the day/night terminator and any thermal inertia, leading to a prolonged activity in previously illuminated areas rotating into the night hemisphere.
For the demonstration of the collimation effect and the bending of jets within 20~km from the nucleus, we do not take into account that during the dust emission new surface areas get illuminated and contribute to the column density, since the dust spends about 1~h (Fig.~\ref{fig:forces}) within the 20~km zone around the nucleus, which is significantly less compared to the cometary rotation period of 12~h.\\
If the dust velocity near the surface is small, the dust trajectories bend strongly due to the Coriolis effect \cite{Kramer2015,Kramer2015b}.
This is the case for a reduced gas-dust interaction (for instance by a reduced gas density in the shaded areas) and seen in observation and simulation for jets originating from the shaded surface areas (Fig.~\ref{fig:night}).

\subsection{Dust column densities}

To compare the dust model with Rosetta observations of 67P/C-G we simulate the dust column density along the camera direction extending $\pm 20$~km from the center of the nucleus.
The dust column density is obtained by binning the positions at all contributing trajectories into a grid with cell size $0.4\times0.4\times40$~km$^3$, where the long axis is aligned with the viewing direction of the OSIRIS camera.
The finite number of representative dust particles results in streaks, which become more apparent with increasing distance from the nucleus.
With a higher resolution shape model and the assumption of a homogeneous gas emission, a 10 times better resolution (50~m) is achieved \cite{Kramer2015,Kramer2015b,Kramer2016}.
Here, we focus on the inclusion of the best-fit gas model derived directly from the COPS measurements, which limits the number of representative dust emitters to 9996 (one dust emitter per surface cell and gas source).

\subsection{Coriolis effect}\label{sec:coriolis}

\begin{figure}[t]
\begin{center}
\includegraphics[width=\textwidth]{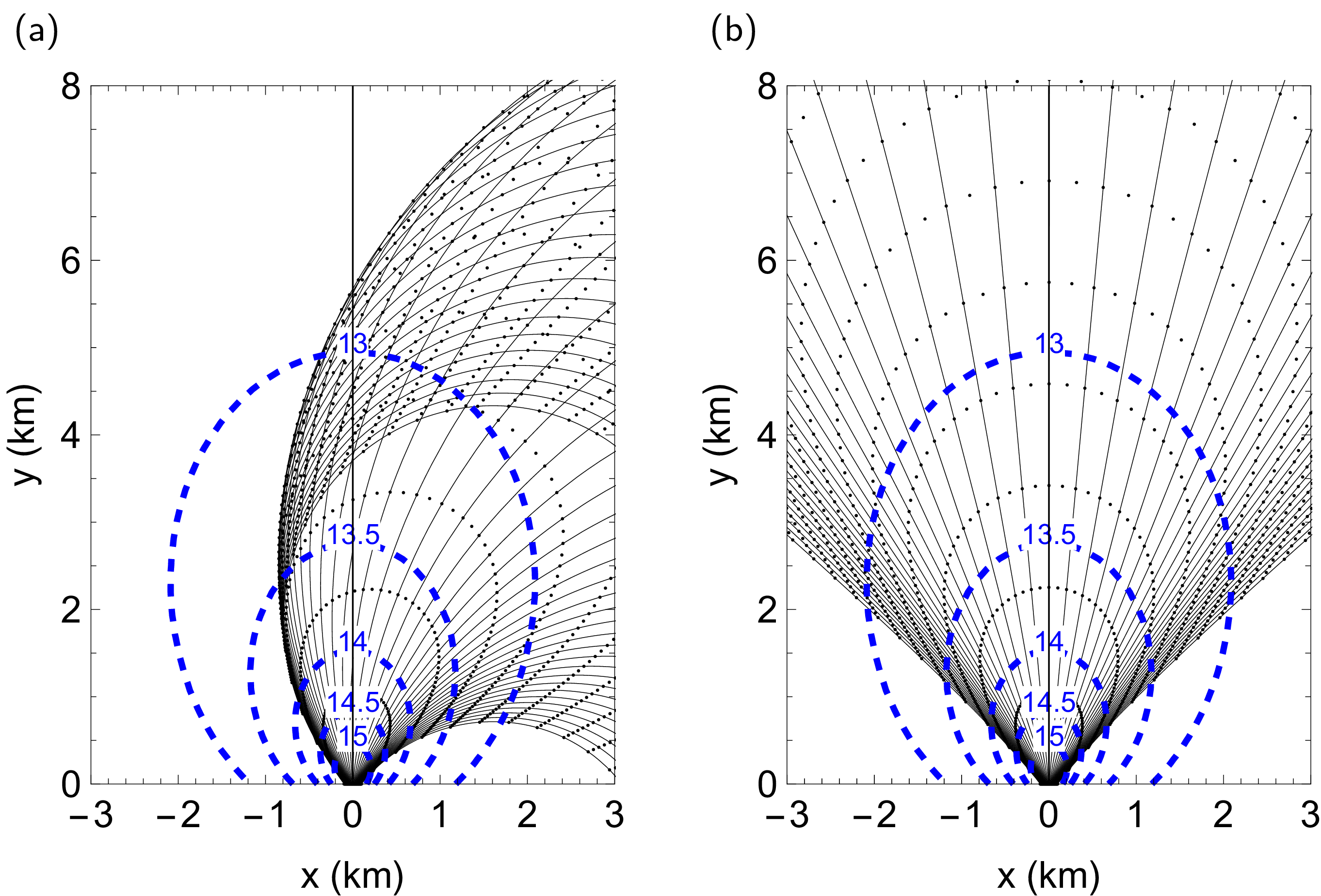}
\end{center}
\caption{\label{fig:singlejet} 
Dust trajectories (black lines) accelerated within a single gas emitting source
(dashed blue lines: $\log_{10}(N_\text{gas}(x,y))={\rm const}$) for $\alpha=8.5$, 
$N_\text{gas}(0,0)=5\times 10^{17}$~H$_2$O molecules/m$^3$, $v_\text{gas}=540$~m/s.
The initial ground locations of the dust particles is distributed along a line $x=-0.2,\ldots,0.2$~km.
The curved dust flow is caused by the Coriolis effect of the rotating nucleus, shown here at (a) maximal deflection and (b) without taking rotation into account.
}
\end{figure}

\begin{figure}[t]
\begin{center}
\includegraphics[width=0.99\textwidth]{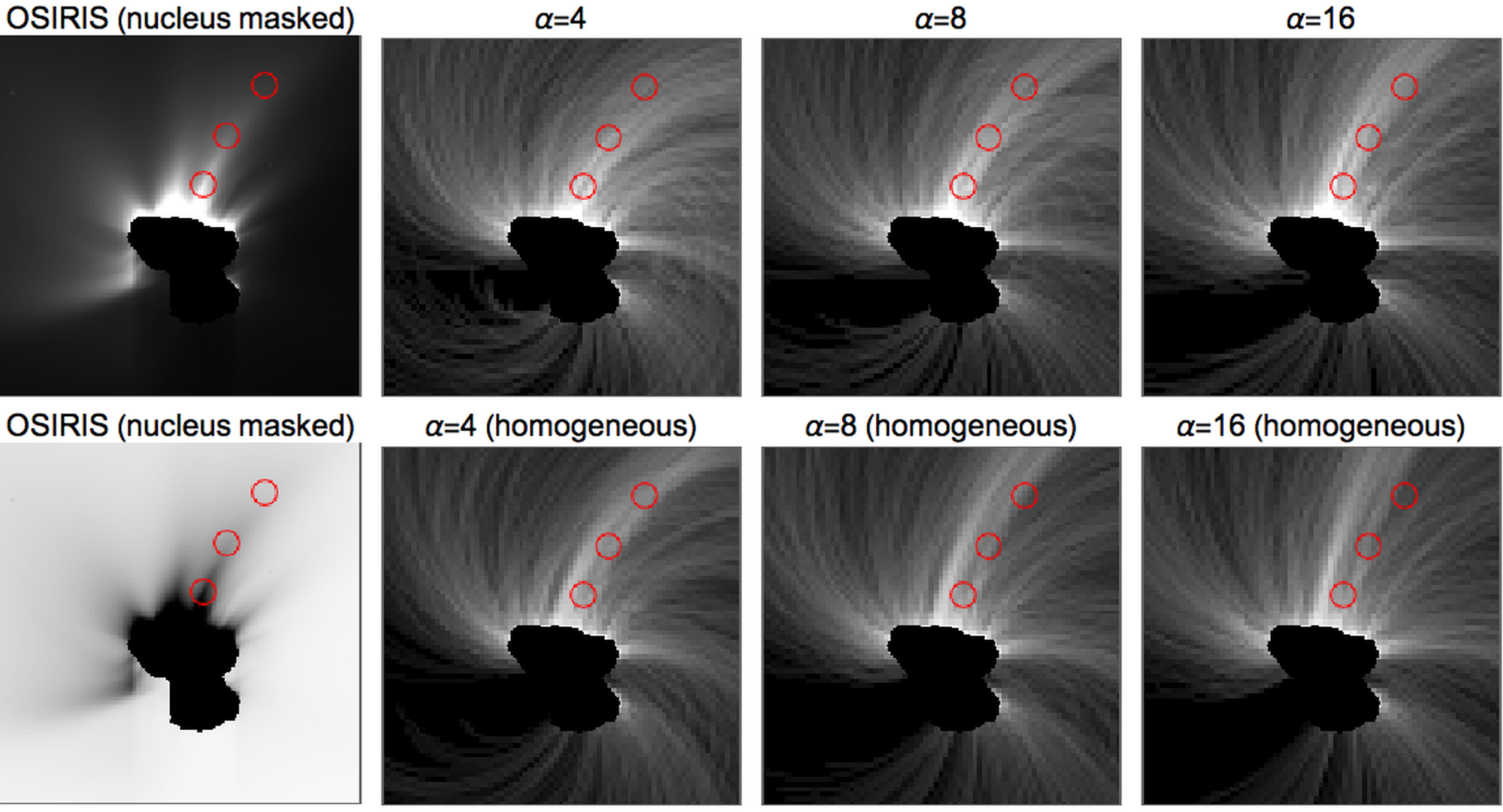}
\end{center}
\caption{\label{fig:Apr22}
Parametric study of varying dust-gas interactions ($\alpha= 4,8,16$) for Rosetta's position on 2014-04-22 21:08 UTC, 137~km away from the center of the nucleus.
The red circles, at the same positions in all panels, mark the location of the extended, curved jet in the reference OSIRIS WAC image (id: W20150422T210738516ID30F18, filters: empty+VIS610, exposure: 7.8~s, direction to Sun upwards).
Best agreement of observation and simulation is observed between $\alpha=8$ and $16$ in combination with the best-fit gas coma (upper row).
The homogeneous gas emitter solution accelerates dust more effectively and agrees best for $\alpha=4$.
}
\end{figure}

The Coriolis effect is exemplified for a single gas emitting source in Fig.~\ref{fig:singlejet}, which shows that localized gas sources lead to more curved dust trajectories.
On the other hand, a homogeneous distribution of gas emitters on the surface leads to a further acceleration of deflected particles and to a longer lasting acceleration, as seen in Fig.~\ref{fig:Apr22} by the decreased jet curvature in the lower panel (homogeneous distribution of gas sources, corresponding gas coma depicted in Fig.~\ref{fig:gasdensitymap}b) for the same value of $\alpha$.

Curved trajectories are clearly discernible in the OSIRIS image from Apr 22, 2015, see Fig.~\ref{fig:Apr22}.
A comparison with computed dust trajectories leads to an estimate of the velocity $3-4$~m/s by the curvature of the trajectory (Eq.~(\ref{eq:rcor})), in line with GIADA measurements at  $r\sim$~25~km \cite{Fulle2015} up to Jan 2015.
The Coriolis effect puts strong constraints on the initial acceleration. 
A slow onset of dust-gas interactions ($\alpha=4$, Fig.~\ref{fig:Apr22}) leads to a large sidewards drift due to the Coriolis force and 
redistributes material on the comet by back fall \cite{Kramer2015,Kramer2015b}. 
With an increase in gas-dust interaction a faster velocity gain is possible and for $\alpha=16$ the observed bending closely resembles the simulation ($\alpha=16$, Fig.~\ref{fig:Apr22}). 
Due to the high gas velocity ($540$~m/s) compared to the dust velocity ($< 10$~m/s), the bending of the gas flux due to the rotation of the nucleus is negligible in the innermost coma ($<20$~km).

\section{Results}\label{sec:results}

\begin{figure}[t]
\begin{center}
\includegraphics[width=\textwidth]{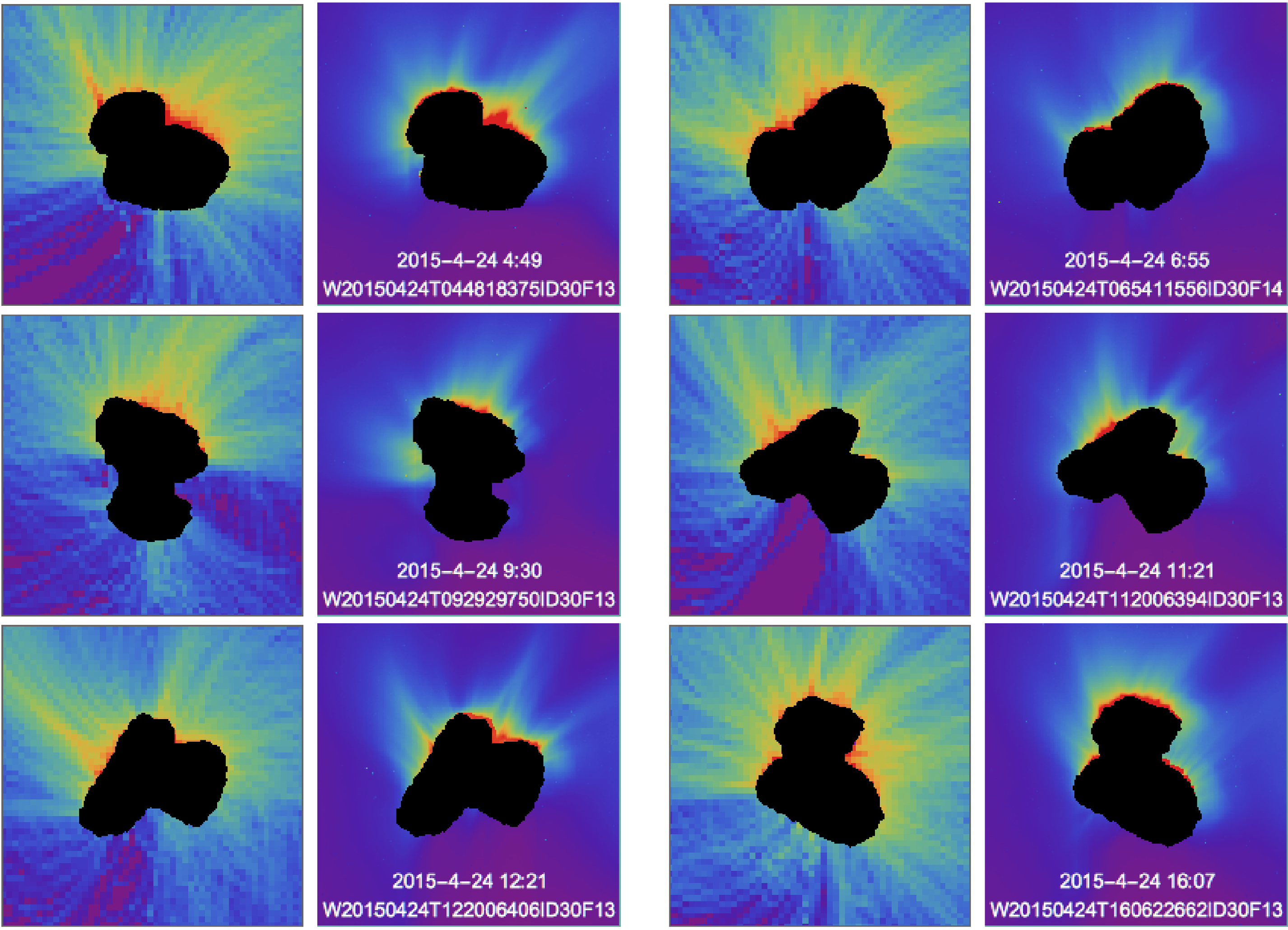}
\end{center}
\caption{\label{fig:Apr24} 
Dust coma of 67P/C-G on 2015-04-24 from a distance of 100~km.
Left panels: simulated dust column density ($\alpha=8$) during a diurnal rotation period, right panels: OSIRIS WAC image (ids given in insets, filters: Empty+UV375, exposure 36.45~s). Sun direction is pointing upwards, the shadow of the nucleus on the emitted dust particles is not considered in the simulation.
}
\end{figure}

The computed results show fine-structured and collimated jets within the dust distribution and the column dust density during a full diurnal period (Fig.~\ref{fig:Apr24}) with changing viewing geometry and illumination.
No model parameters are adjusted to compare the results across differing rotation periods. 
The dust jets are forming despite the homogeneous dust release across the entire (sunlit) surface and are naturally arising from the simulation, without the need for assigning isolated patches of activity \cite{Kramer2016}.
The collimation is driven by the cometary surface topography and reflects concave (crater-like and larger scale) surface features, whose surface normals intersect in a focal area several km away from the surface.
Close to the nucleus, a correlation up to $0.9$ (average $0.8$) between a homogeneous dust emission scenario and the observed OSIRIS image intensities has been established during a complete diurnal rotation period \cite{Kramer2016}, including recurring ray-type structures caused by the alignment of multiple concavities along the line of sight.
As shown here, the inclusion of a best-fit gas-coma model does not only preserve the collimation, but in addition allows one to study the bending of dust trajectories further away from the nucleus.
The collimation and bundling of trajectories, which are initially more uniformly distributed either in position or momentum space, is a generic physical feature 
seen whenever a perturbation slightly distorts an otherwise uniformly propagating trajectory field.
Similar effects related to ray caustics and lensing of trajectories were observed on the nanometer scale leading to branched electron flow over a random potential \cite{Topinka2001a} and on the kilometers scale for freak waves in otherwise almost Gaussian seas \cite{Heller2008}.

\section{Conclusions}

The Rosetta mission has returned a wealth of observational data about the evolution of a comet on its orbit around the sun.
To bring the dust and gas emission models in agreement with measurements requires considering, in particular, the concave shape features of the nucleus.
We have shown that homogeneous gas and dust emission on a detailed topographic model of the rotating cometary nucleus accounts for many jet-like features seen in the dust coma observations of 67P/C-G.
The good agreement between simulated dust coma and observations   separated by several cometary rotation periods strengthens the hypothesis of a largely static and homogeneous gas and dust emission scenario. 
Additional sporadic (5~min) outbreaks of dust jets \cite{Vincent2016} are not considered in the predictive dust coma model, but require to resolve the heterogeneity of the surface composition.
The Coriolis effect explains the curvature of the jets seen in the OSIRIS images of 67P/C-G and allows one to determine the dust velocity within specific jets ($3$-$4$~m/s).
Further improvements of cometary dust emission models are required to understand in even more detail the Rosetta data and the activities of other comets.
Isolated regions of reduced activity and day/night effects could be readily incorporated in the model by removing dust trajectories originating from the corresponding surface cells.
Such a fully time-dependent simulation of dust and gas emission profiles might increase the agreement in the night and terminator region.
In addition the seasonal and shifting gas activity \cite{Kramer2017}, with maximum activity occurring 20 days after perihelion (southern solstice), induces material transport and back fall of larger dust particles on the surface.  
The complex cometary shapes seen at several different spacecraft encounters call for efficient simulation and visualization algorithms.
Already for the first observed periodic comet 1P/Halley, the overall bean-like concave shape has been proposed as being responsible for the larger concentration of dust in its focal direction \cite{Crifo2002}.

The driving forces leading to a bilobed shape are not yet understood and could point to a binary object formation or a runaway process during the orbital evolution.
A related open question concerns the precise microscopic release conditions of the gas and dust at and below the surface.
In particular the simulations shown here point to a fast initial acceleration of the dust, which requires either spatially confined gas sources, albeit distributed across the entire nucleus, or might indicate that other forces besides gas and gravitation play a role (e.g. surface tensile strength and gas pressure build up, or electric charging) \cite{Weidling2012,Nordheim2015}.

\enlargethispage{2cm}

\section*{Acknowledgements}

We thank Mattias~Malmer (Stockholm, Sweden) for helpful discussions and providing the shape model of comet 67P/C-G. We are indebted to Kathrin Altwegg and Martin Rubin from the ESA/Rosetta ROSINA team (University of Bern, Switzerland) for helpful discussions.
TK acknowledges support by a Heisenberg fellowship of the DFG and helpful discussions with Peter~Kramer (University of T\"ubingen, Germany).
Computational resources and support by the North-German Supercomputing Alliance (HLRN) are gratefully acknowledged. 
The authors acknowledge the OSIRIS Principal Investigator Holger~Sierks (MPS, G\"ottingen, Germany) and the OSIRIS Team for providing images and related datasets and the ESA Rosetta Project to enable the science of the mission.
The shown OSIRIS images were retrieved from the ESA Planetary Science Archive and NASA Planetary Data System (2016), curated by  P.~Gutierrez-Marques, H.~Sierks and the OSIRIS Team, data set identifier: ROSETTA-ORBITER COMET ESCORT OSIWAC 3 RDR MTP 015 V1.0, RO-C-OSIWAC-3-ESC2-67PCHURYUMOV-M15-V1.0.

\bibliographystyle{tADP}

\clearpage

\end{document}